\documentclass[conference]{IEEEtran}
\IEEEoverridecommandlockouts
\usepackage{cite}
\usepackage{amsmath,amssymb,amsfonts}
\usepackage{subcaption}
\usepackage{algorithmic}
\usepackage{graphicx}
\usepackage{textcomp}
\usepackage{xcolor}
\usepackage{physics}
\usepackage{multirow}
\usepackage{bm}
\usepackage{array}
\usepackage[a4paper, total={184mm,239mm}]{geometry}
\def\BibTeX{{\rm B\kern-.05em{\sc i\kern-.025em b}\kern-.08em
    T\kern-.1667em\lower.7ex\hbox{E}\kern-.125emX}}
\begin{document}
\title{How Parallel Circuit Execution Can Be Useful for NISQ Computing?\\

}

\author{\IEEEauthorblockN{Siyuan Niu}
\IEEEauthorblockA{\textit{LIRMM, University of Montpellier} \\
	34090, Montpellier, France \\
	siyuan.niu@lirmm.fr}
\and
\IEEEauthorblockN{Aida Todri-Sanial}
\IEEEauthorblockA{\textit{LIRMM, University of Montpellier, CNRS} \\
	34090, Montpellier, France \\
	aida.todri@lirmm.fr}
}

\maketitle

\begin{abstract}
Quantum computing is performed on Noisy Intermediate-Scale Quantum (NISQ) hardware in the short term. Only small circuits can be executed reliably on a quantum machine due to the unavoidable noisy quantum operations on NISQ devices, leading to the under-utilization of hardware resources. With the growing demand to access quantum hardware, how to utilize it more efficiently while maintaining output fidelity is becoming a timely issue. A parallel circuit execution technique has been proposed to address this problem by executing multiple programs on hardware simultaneously. It can improve the hardware throughput and reduce the overall runtime. However, accumulative noises such as crosstalk can decrease the output fidelity in parallel workload execution. In this paper, we first give an in-depth overview of state-of-the-art parallel circuit execution methods. Second, we propose a Quantum Crosstalk-aware Parallel workload execution method (QuCP) without the overhead of crosstalk characterization. Third, we investigate the trade-off between hardware throughput and fidelity loss to explore the hardware limitation with parallel circuit execution. Finally, we apply parallel circuit execution to VQE and zero-noise extrapolation error mitigation method to showcase its various applications on advancing NISQ computing.

\end{abstract}

\section{Introduction}
Today's quantum computers are qualified as Noisy Intermediate-Scale Quantum (NISQ) devices with 50 to hundreds of qubits~\cite{preskill2018quantum}. It is limited by several physical constraints and noisy quantum operations. There are not enough qubits to realize the quantum error correction codes (QECC)~\cite{calderbank1998quantum} for a universal fault-tolerant quantum computer. Current quantum chips give reliable results only when executing a small circuit with shallow depth, causing a waste of hardware resources. Moreover, there is a growing demand to access quantum devices via the cloud, which leads to a large number of jobs in the queue and long waiting times for users. For example, it takes several days to get the result if we submit a circuit on IBM public quantum chips. Therefore, how to efficiently make use of quantum hardware to reduce the total runtime of circuits is becoming a timely problem. 

The parallel circuit execution technique was firstly proposed by~\cite{das2019case} to target this problem. It allows a user to execute several quantum programs on a quantum chip simultaneously, or multiple users can share one quantum device at the same time. It improves the quantum hardware throughput and reduces the users' waiting time. But the results show that the output fidelities of these circuits are decreased. Other approaches~\cite{liuqucloud, niu2021enabling, ohkuracrosstalk} have been proposed to enhance this technique, introducing different circuit partition methods, mapping algorithms, and taking crosstalk into account. Their results demonstrate that parallel circuit execution can be particularly of interest to quantum applications requiring simultaneous sub-problem executions. 

In this paper, we focus on investigating how parallel circuit execution can be useful for NISQ computing. Our major contributions can be listed as follows:
\begin{itemize}
\item We provide an in-depth overview of parallel workload executions and outline the advantages and limitations.

\item We propose a Quantum Crosstalk-aware Parallel workload execution method (QuCP) which considers crosstalk error while eliminating the significant overhead of crosstalk characterization methods.

\item We perform parallel circuit execution on IBM quantum devices and analyze the hardware limitation of executing multiple circuits simultaneously. 

\item We apply parallel circuit execution to VQE and zero-noise extrapolation (ZNE) to demonstrate its applications on NISQ algorithms and error mitigation techniques. 

\end{itemize}

\section{Background and State of the Art}

\subsection{Introduction of Parallel Circuit Execution}
As the size of a quantum chip increases, there is a need to execute multiple shallow depth circuits in parallel. This improves not only hardware throughput (the number of used qubits divided by the total number of qubits) but also reduces the overall runtime (waiting time + execution time).

\begin{figure}[h]
\begin{subfigure}{\linewidth}
\centering
\includegraphics{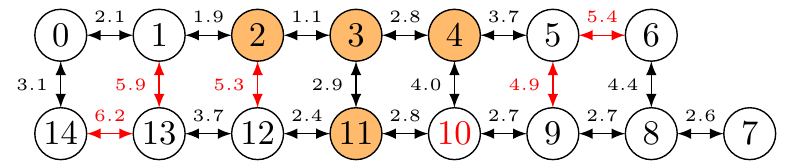}
\caption{Execute one circuit with hardware utilization of 26.7\%}
\label{fig:multi_1}
\end{subfigure}\\
\begin{subfigure}{\linewidth}
\centering
\includegraphics{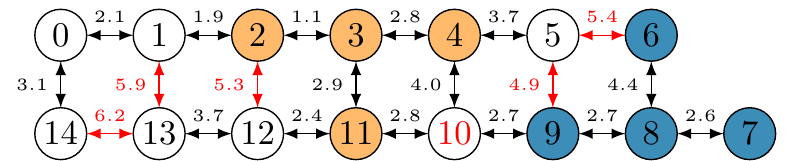}
\caption{Execute two circuits with hardware utilization of 53.3\%}
\label{fig:multi_2}
\end{subfigure}
\caption{An example of parallel circuit execution on IBM Q 16 Melbourne. The links with high \texttt{CNOT} error rate and the qubits with high readout error rate are marked in red color. One-qubit error is ignored because of its relatively low error rate.}
\label{fig:multiprogramming}
\end{figure}

Fig.~\ref{fig:multi_1} shows an example of executing one 4-qubit quantum circuit on IBM Q 16 Melbourne. The circuit is mapped to a reliable region with tense-connectivity. In this case, the hardware throughput is only 26.7\% and most of the qubits are unused. It is possible to find another reliable region to run two 4-qubit circuits in parallel, as shown in Fig.~\ref{fig:multi_2}. The hardware throughput is increased to 53.3\%, and the total runtime is reduced by half.

However, as hardware throughput increases, output fidelity is reduced because: (1) Qubits with high fidelities are sparsely distributed, and it is difficult to execute all the quantum circuits on reliable regions. (2) Running multiple circuits in parallel can introduce a higher chance of crosstalk error~\cite{sheldon2016procedure}. How to trade off the output fidelity and hardware throughput to benchmark the IBM quantum hardware limitation is the focus of our work and discussed in section~\ref{section:experiment}. 

\subsection{Review and Comparison with State of the Art}
There are mainly two steps to realize a parallel circuit execution method: (1) Allocate partitions to multiple circuits and make sure they do not interact with each other during execution. (2) Make all the circuits executable on hardware using parallel qubit mapping approach. Here, we compare the state-of-the-art methods: MultiQC~\cite{das2019case}, QuCloud~\cite{liuqucloud}, QuMC~\cite{niu2021enabling}, and CNA~\cite{ohkuracrosstalk}, and discuss the key features to design a parallel workload execution algorithm.

\textbf{Crosstalk.} It is one of the major noise sources in NISQ devices and can reduce the output fidelity significantly~\cite{murali2020software}. 
When multiple quantum operations are executed simultaneously, the state of one qubit might be corrupted by the operations on the other qubits if crosstalk exists. In parallel circuit execution, as several circuits are executed simultaneously, the probability of crosstalk is increased. Crosstalk is considered at partition-level to be avoided between partitions in QuMC, whereas CNA considers it at gate-level during the qubit mapping process. 

\textbf{Characterization of crosstalk.} In order to consider crosstalk in parallel circuit execution, we must figure out how to characterize it. Both QuMC and CNA use Simultaneous Randomized Benchmarking (SRB)~\cite{gambetta2012characterization} to characterize crosstalk of the target quantum device because of its ability to quantify the crosstalk impact between simultaneous \texttt{CNOT} operations. However, this approach becomes expensive with the increase in the size of a quantum chip. Better crosstalk characterization or crosstalk mitigation method is needed.

\textbf{Qubit partitioning.} This process aims to allocate reliable partitions to each program. Except CNA, all the previous works propose their qubit partition algorithms, taking hardware topology and calibration data into account. In addition, QuMC considers crosstalk during qubit partition. 

\textbf{Qubit mapping.} The objective is to make quantum circuits executable on quantum hardware regarding the hardware topology. It includes two parts: initial mapping and routing. MultiQC and QuMC use a noise-aware mapping approach~\cite{niu2020hardware}, whereas CNA chooses another noise-adaptive method~\cite{murali2019noise} while considering crosstalk. QuCloud considers both inter and intra-program \texttt{SWAP}s to reduce the \texttt{SWAP} number but introduces potential crosstalk error.

\textbf{Task scheduling.} All these methods use As Late As Possible (ALAP) approach for task scheduling, allowing qubits to remain in the ground state for as long as possible. It avoids the extra decoherence error caused by parallel execution of circuits with different depths and is the default scheduling method used in Qiskit compiler~\cite{Qiskit-Textbook}. 

\textbf{Independent vs Correlated.} One important question in parallel circuit execution is to determine the number of circuits to execute simultaneously. QuCloud and QuMC propose different metrics to estimate the fidelity of allocated partition. QuMC further introduces a fidelity threshold to select the optimal number of simultaneous circuits.  

In conclusion, QuMC covers all the important factors in designing a parallel circuit execution method and has reported the best results in their paper compared with MultiQC and QuCloud. However, it still has the drawback of the large overhead when performing SRB for crosstalk characterization, which limits its performance when applied to large-scale quantum devices. It is essential to address this problem because the parallel circuit execution technique is especially dedicated to small benchmarks on large machines. 

\textbf{}

%

\section{Quantum Crosstalk-aware Parallel Execution}

To address the drawbacks of the previous works, we propose a Quantum Crosstalk-aware Parallel workload execution method (QuCP) which emulates the crosstalk impact without the overhead of characterizing it. 

Simultaneous Randomized Benchmarking is the most popular approach to quantify the crosstalk properties of a quantum device. For example, suppose we want to characterize the crosstalk effect between one pair of two \texttt{CNOTs} $g_i$ and $g_j$. In that case, we need to first perform Randomized Benchmarking (RB) on the two \texttt{CNOTs} individually and then make simultaneous RB sequences on this pair. This process introduces a significant overhead if applied to large devices. Crosstalk is shown to be significant between neighbor \texttt{CNOT} pairs and~\cite{murali2020software} proposed several optimization methods to lower SRB overhead by grouping \texttt{CNOT} pairs separated by more than one-hop distance and performing SRB on them simultaneously. However, SRB is still expensive even with these optimization methods. The overhead of performing SRB on two quantum chips: IBM Q 27 Toronto and IBM Q 65 Manhattan, is shown in Table~\ref{table1}. The one-hop pairs (neighbor \texttt{CNOT} pairs) are allocated to a minimum number of groups. We choose 5 seeds to ensure the precise result of SRB, and the number of jobs needed to perform SRB is 135 and 165, respectively, which takes a significant amount of time. The cost becomes even worse as the size of the quantum chip increases. Despite the expensive cost, SRB also requires users to master this technique to characterize crosstalk, which is not trivial. 

Inspired by QuMC, which mitigates crosstalk error at partition-level, we introduce a crosstalk parameter $\sigma$ to represent the crosstalk impact on \texttt{CNOT} pairs without the need of learning and performing SRB. Given a list of circuits to execute simultaneously, we first use the heuristic qubit partitioning method from QuMC to allocate the partition for the first circuit and add these qubits to a list of allocated qubits $q_{allocate}$. For the rest of the circuits, each time when we construct the possible partition candidates, we check if there are some pairs inside of the partition candidate are a one-hop distance from the pairs inside of $q_{allocate}$ according to the hardware topology. If there exists some, we can collect a list of potential crosstalk pairs $q_{crosstalk}$. To select the best partition, we calculate the Estimated Fidelity Score ($EFS$) of all the partition candidates shown in~\eqref{EFS}. 

\begin{equation}
	EFS = Avg_{2q(cross)} \times \#2q + Avg_{1q} \times \#1q + \sum_{Q_i \in P} R_{Q_i}
\label{EFS}
\end{equation}

$Avg_{2q(cross)}$ is the average 2-qubit (\texttt{CNOT}) error inside of the partition $P$. Note that if $q_{crosstalk}$ is not empty, we use the crosstalk parameter $\sigma$ times the \texttt{CNOT} errors of the pairs inside of $q_{crosstalk}$ to indicate the crosstalk effect. Similarly, $Avg_{1q}$ is the average 1-qubit error rate, and $R_{Q_i}$ is the readout error of the qubit $Q_i$ belonging to partition $P$. We can use this metric $EFS$ to emulate the impact of crosstalk and avoid it at partition-level without performing SRB to characterize the crosstalk properties of a quantum device.

\begin{table}
\caption{Overhead of SRB on different IBM quantum chips}
\label{table1}
\begin{center}
\begin{tabular}{|l|l|l|}
	\hline
	\textbf{Chip} & \textbf{IBM Q 27 Toronto} & \textbf{IBM Q 65 Manhattan} \\
	\hline
	qubit & 27 & 65\\
	\hline
	1-hop pairs & 28 & 72\\
	\hline
	groups & 9 & 11\\
	\hline
	seeds & 5 & 5\\
	\hline
	jobs & 135 & 165\\
	\hline
\end{tabular}
\end{center}
\end{table}

\section{Experimental results}
In this section, we first evaluate the performance of our QuCP method by comparing it with state-of-the-art crosstalk-aware parallel circuit execution algorithms. Second, we explore the hardware limitation when performing parallel circuit execution. Finally, we demonstrate the benefit of applying parallel circuit execution to VQE and ZNE algorithms.

\begin{figure}[t]
\centering
\includegraphics[scale=0.3]{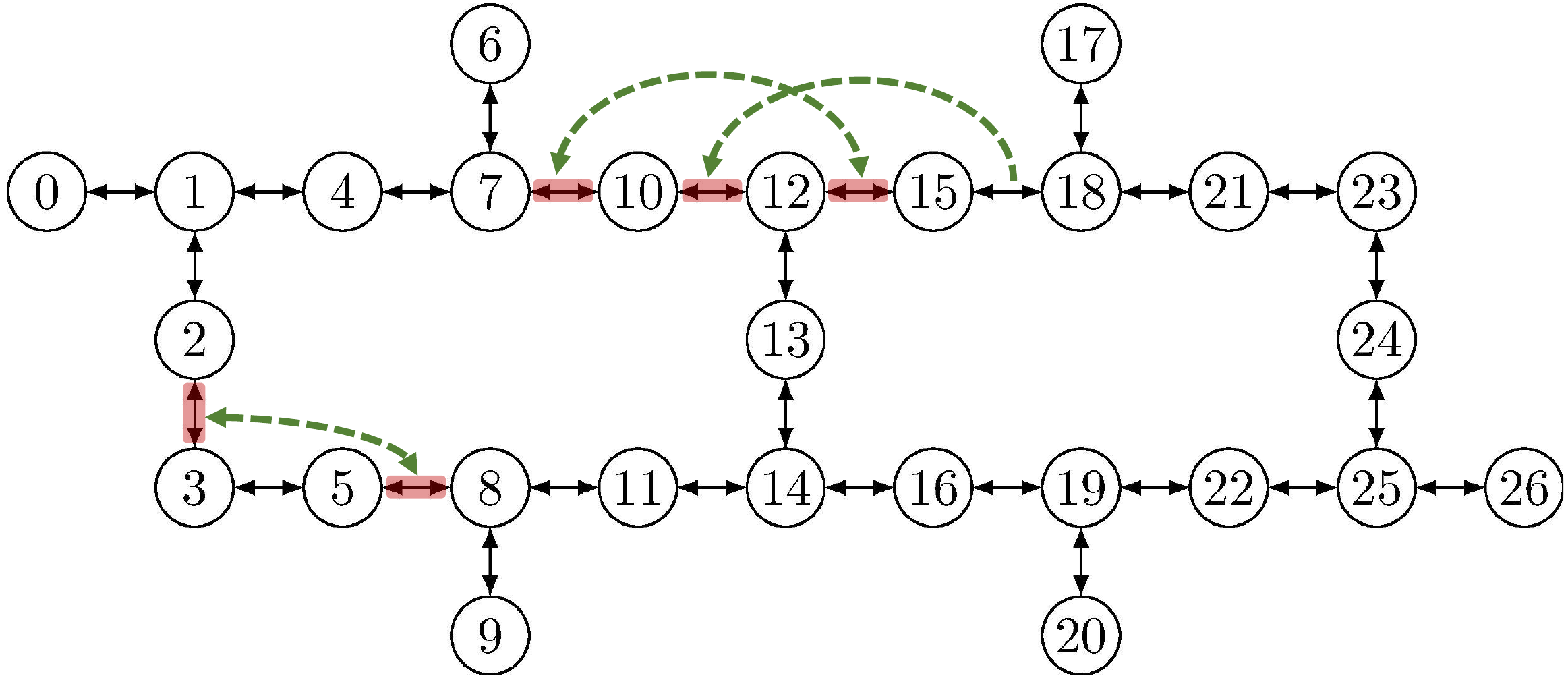}
\caption{Crosstalk characterization of IBM Q 27 Toronto using Simultaneous Randomized Benchmarking. The \texttt{CNOT} pairs that are significantly influenced by crosstalk are highlighted by arrows and red color.}
\label{fig:toronto_crosstalk}
\end{figure}

\subsection{Crosstalk-aware Parallel Circuit Execution}

\begin{figure}[h]
\begin{subfigure}{0.45\linewidth}
\centering
\includegraphics[scale=0.4]{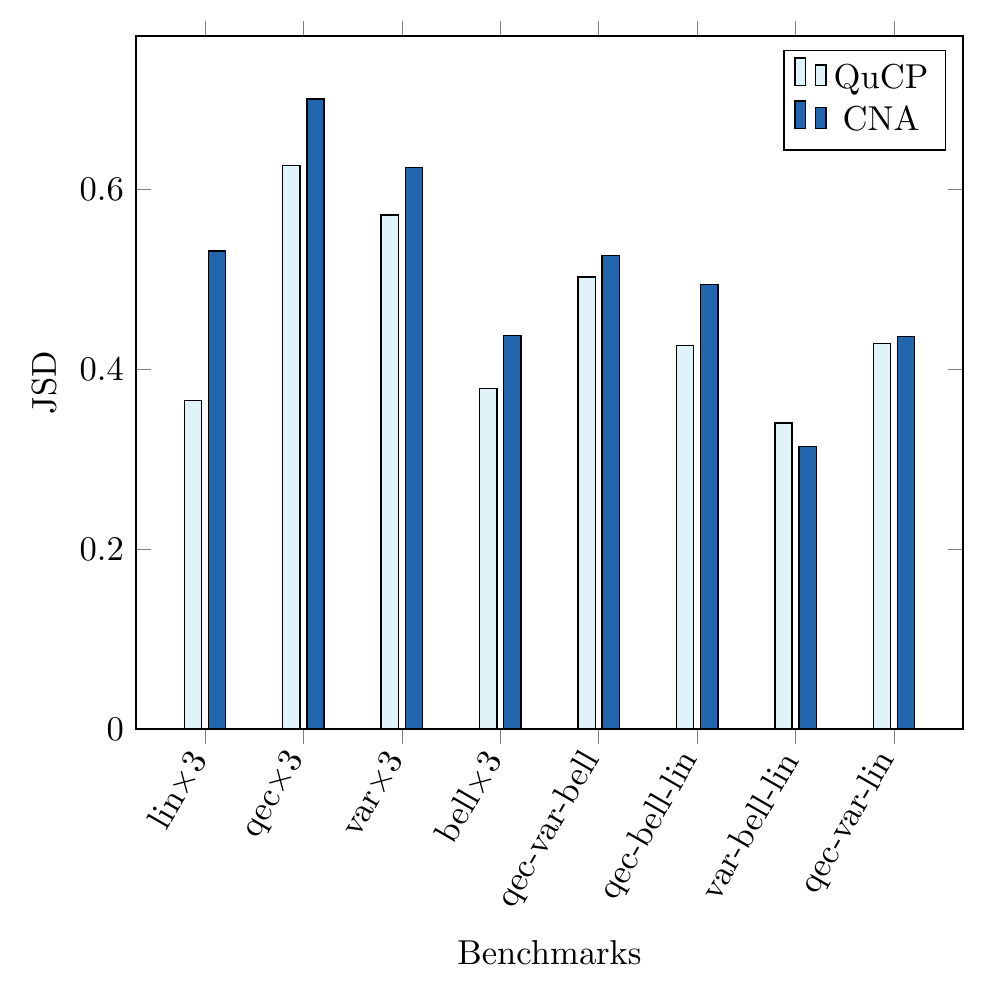}
\caption{JSD result (lower is better).}
\label{fig:jsd}
\end{subfigure}
\begin{subfigure}{0.45\linewidth}
\centering
\includegraphics[scale=0.4]{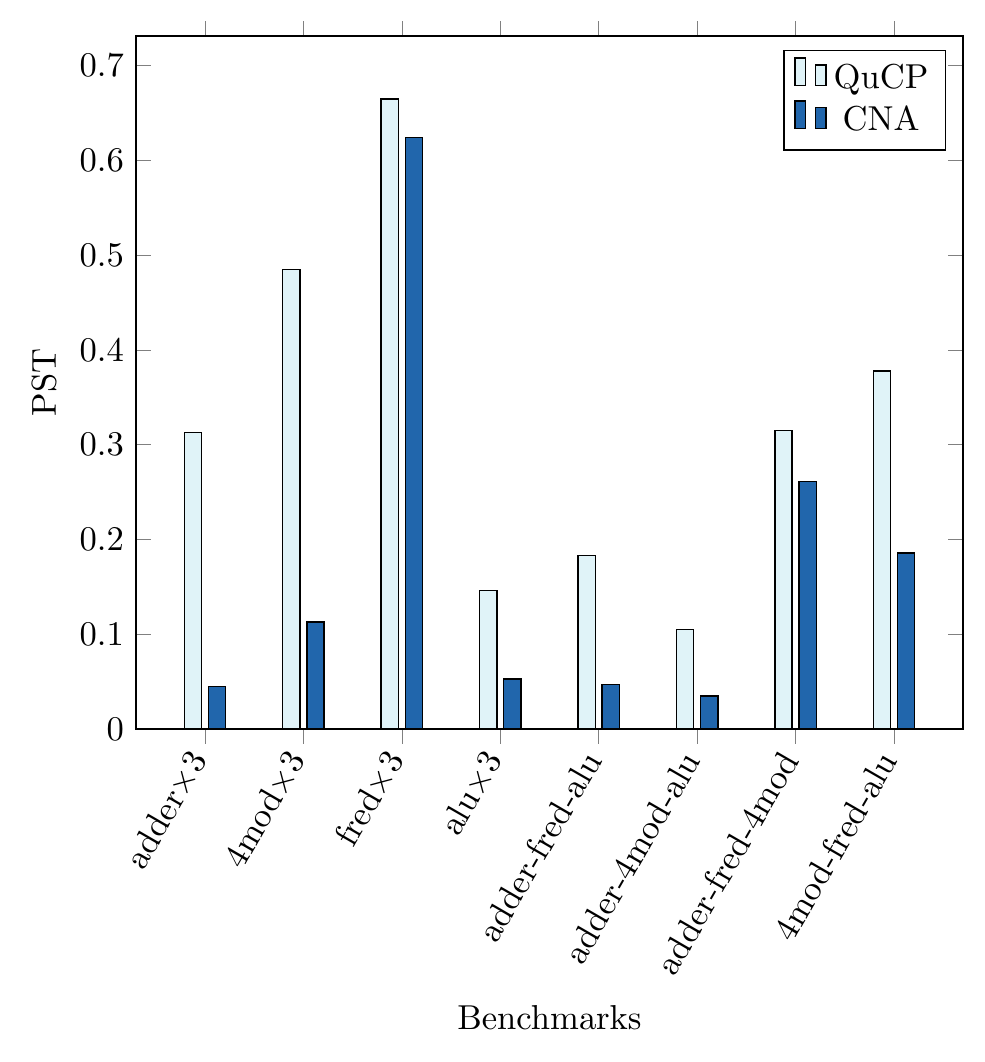}
\caption{PST result (higher is better).}
\label{fig:pst}
\end{subfigure}
\caption{The fidelity result of executing three benchmarks simultaneously on IBM Q 27 in terms of JSD and PST metrics. The experiments include combining three different benchmarks and repeating the same benchmark three times. }
\label{fig:crosstalk_algorithm}

\end{figure}

We compare our QuCP with two crosstalk-aware parallel workload execution approaches, QuMC and CNA, which are different in terms of crosstalk-mitigation method, qubit partitioning, and qubit mapping process. Both of the two methods need to perform SRB for crosstalk characterization. 

We use SRB to characterize the crosstalk properties of IBM Q 27 Toronto (Fig.~\ref{fig:toronto_crosstalk}). Table~\ref{table2} shows the benchmarks that we use to compare these algorithms. They are collected from~\cite{li2020qasmbench,wille2008revlib}, including several functions about logical operations, error correction, and quantum simulation, etc.

We calculate the output fidelity of the simultaneous circuits to evaluate the performance of these algorithms. Some of the benchmarks have one certain output, and we use the Probability of a Successful Trial (PST) metric defined in~\eqref{eq:1}. Whereas for other benchmarks, their results are supposed to be a distribution. We choose Jensen-Shanno Divergence (JSD) to compare the distance of two probability distributions, shown in~\eqref{eq:2}, where $P$ and $Q$ are two distributions to compare and $M = \frac{1}{2}(P + Q)$. It is based on Kullback-Leibler divergence, shown in~\eqref{eq:3}, with the benefit of always having a finite value and being symmetric. 

\begin{equation}
PST = \frac{\text{Number of successful trials}}{\text{Total number of trials}}
\label{eq:1}
\end{equation}

\begin{equation}
JSD(P || Q) = \frac{1}{2}D(P||M) + \frac{1}{2}D(Q || M)
\label{eq:2}
\end{equation}

\begin{equation}
D_{KL}(P||Q) = \sum_{x \in \mathcal{X}}P(x)\log(\frac{P(x)}{Q(x)})
\label{eq:3}
\end{equation}

\begin {table}[t]
\begin{center}
\caption{Information of benchmarks}
\label{table2}
\begin{tabular}{|l|l|l|l|l|}
\hline
\textbf{Benchmark} & \textbf{Qubits} & \textbf{Gates} & \textbf{CX} & \textbf{Result} \\
\hline
adder & 4 & 23 & 10 & 1\\
\hline
linearsolver & 3 & 19 & 4 & dist\\
\hline
4mod5-v1\_22 & 5 & 21 & 11 & 1\\
\hline
fredkin & 3 & 19 & 8 & 1\\
\hline
qec\_en & 5 & 25 & 10 & dist\\
\hline
alu-v0\_27 & 5 & 36 & 17 & 1\\
\hline
bell & 4 & 33 & 7 & dist\\
\hline
variation & 4 & 54 & 16 & dist\\
\hline
\end{tabular}
\end{center}
\end{table}

We execute three benchmarks on IBM Q 27 Toronto in parallel. The \texttt{optimization\_level} in Qiskit compiler is set to 3, which is the highest level for circuit optimizations. 

First, we tune the crosstalk parameter $\sigma$ used in QuCP to verify its ability for crosstalk-mitigation at partition-level without SRB by comparing its partitioning results with QuMC. When $\sigma \geq 4$, QuCP provides the same results as QuMC. This number is reasonable as we need to calculate the average \texttt{CNOT} error rate inside of the partition (see~\eqref{EFS}), which can decrease the impact of crosstalk on \texttt{CNOT} pair. Based on this experiment, we set $\sigma$ to 4 and compare QuCP with CNA to show the influence of crosstalk-mitigation at partition-level or gate-level for parallel circuit execution. 

The results in terms of JSD and PST are shown in Fig.~\ref{fig:crosstalk_algorithm}. Note that a lower JSD or a higher PST is desirable. The benchmarks include unitary and various combinations. Comparing QuCP with CNA, the fidelity characterized by JSD and PST is improved by 10.5\% and 89.9\%, respectively. The fidelity improvement is realized by different partitioning and mapping methods, which are two other important factors to consider for parallel circuit execution. QuCP has better results and achieves crosstalk-mitigation with low overhead. 

\subsection{Trade-off Between Hardware Throughput and Output Fidelity}
\label{section:experiment}
\begin{figure}[t]
\begin{subfigure}{\linewidth}
\centering
\includegraphics[scale=0.35]{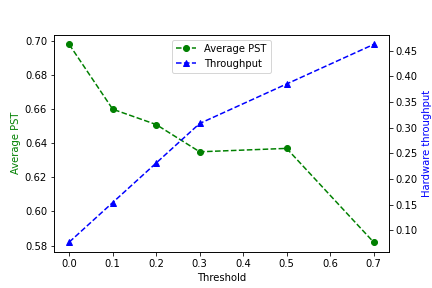}
\caption{\texttt{4mod5-v1\_22} result.}
\label{fig:4mod}
\end{subfigure}
\hspace{0.15cm}
\begin{subfigure}{\linewidth}
\centering
\includegraphics[scale=0.35]{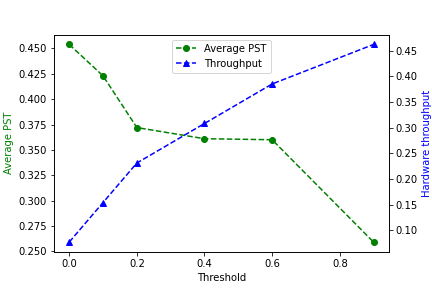}
\caption{\texttt{alu-v0\_27} result.}
\label{fig:alu}
\end{subfigure}
\caption{The result of average PST and hardware throughput with respect to fidelity threshold, which determines the number of simultaneous circuits.}
\label{fig:limitation}

\end{figure}

Enabling parallel circuit execution can improve the hardware throughput significantly. However, it reduces the circuit output fidelity at the same time. It is important to trade-off between them and explore the hardware limitation of performing parallel circuit execution. 

We use our QuCP method to benchmark the hardware limitation. We first estimate the output fidelity difference between independent and parallel circuit executions based on $EFS$ (see~\eqref{eq:1}), then introduce a fidelity threshold to determine how many circuits can be executed simultaneously. Experiments are performed on IBM Q 65 Manhattan, which is IBM's largest quantum chip. We choose two circuits from Table~\ref{table2}: \texttt{4mod5-v1\_22} and \texttt{alu-v0\_27}. We vary the value of the fidelity threshold to execute the same circuit an increasing number of times in parallel, and the results are shown in Fig.~\ref{fig:limitation}.

When the fidelity threshold is zero, which indicates no fidelity difference between independent and simultaneous circuit execution, only one circuit is executed at each time. A larger threshold enables more circuits to be executed simultaneously. The number of parallel circuit executions varies from one to six, corresponding to hardware throughput from 7.7\% to 46.2\% and total runtime reduction up to six times. There is a significant fidelity loss when hardware throughput is over 38\%, which points out the hardware limitation when performing parallel executions for circuits with a similar size as the benchmarks.  

\subsection{Parallel Circuit Execution and VQE Algorithm}
Variational Quantum Eigensolver (VQE)~\cite{kandala2017hardware} is one of the most promising algorithms to achieve quantum advantage in the NISQ era and is recently widely used in quantum chemistry. It can be used to prepare approximations to the ground state energy of a Hamiltonian as a hybrid classical-quantum method with shallow circuits. However, it needs to split the computation into $O(N^4)$ sub-problems, introducing a large overhead of measurement circuits~\cite{gokhale2020optimization}. 

Parallel circuit execution has been used in~\cite{eddins2021doubling} to execute distinct molecular geometries at the same time to increase hardware throughput during VQE routine. Whereas our focus is to investigate parallel circuit execution on independent VQE problem to reduce its measurement overhead. 

A Hamiltonian can be expressed by the sum of tensor products of Pauli operators. For naive measurement, we need quantum circuits for each Pauli term and compute their expectation values to obtain the state energy. This overhead can be reduced by performing simultaneous measurements, grouping commuting Pauli terms to measure them at the same time~\cite{mcclean2016theory, gokhale2020optimization}. Here, we apply parallel circuit execution to VQE to estimate the ground state of molecular \bm{$H_2$} at equilibrium bond length (0.735 angstroms) in the singlet state and with no charge, which can further reduce the measurement overhead. 

We first map the fermionic operators of the Hamiltonian to qubit operators using parity mapping~\cite{bravyi2017tapering} and we obtain a two-qubit Hamiltonian composed of 5 Pauli terms $\{II, IZ, ZI, ZZ, XX\}$. Naive measurements would require one circuit for each Pauli term to calculate the expectation value of the ansatz. These Pauli terms can be partitioned into two commuting groups using simultaneous measurement: $\{II, IZ, ZI, ZZ\}$ and $\{XX\}$. Note that the grouping result is not unique, but two groups are needed.  

\begin {table}[t]
\begin{center}
\caption{Experimental results of the ground state energy of $H_2$ under PG and QuCP+PG with different number of simultaneous measurements ($n_c$ ).}
\label{table:h2}
\begin{tabular}{|l|p{1.2cm}|p{0.4cm}|p{1.3cm}|p{1.3cm}|p{1.8cm}|}
\hline
\multicolumn{2}{|c|}{Experiments} & $n_c$ & $\Delta{E}$\_base (\%) & $\Delta{E}$\_theory & Hardware throughput (\%) \\
\hline
\multirow{2}{*}{(a)} & PG & 1 & 1.4 & 2.6 & 3.1
\\\cline{2-6}
&QuCP+PG & 16 & 2.5 & 3.7 & 49.2 \\
\hline

\multirow{2}{*}{(b)} & PG & 1 & 2.3 & 3.4 & 3.1
\\\cline{2-6}
&QuCP+PG & 20 & 3.8 & 4.9 & 61.5 \\
\hline

\multirow{2}{*}{(c)} & PG & 1 & 1.5 & 2.6 & 3.1 
\\\cline{2-6}
&QuCP+PG & 24 & 5.6 & 6.6 & 73.8 \\
\hline

\end{tabular}
\end{center}
\end{table}

\begin{figure*}[h]
\centering
\begin{subfigure}{0.3\linewidth}
\centering
\caption{}
\includegraphics[scale=0.24]{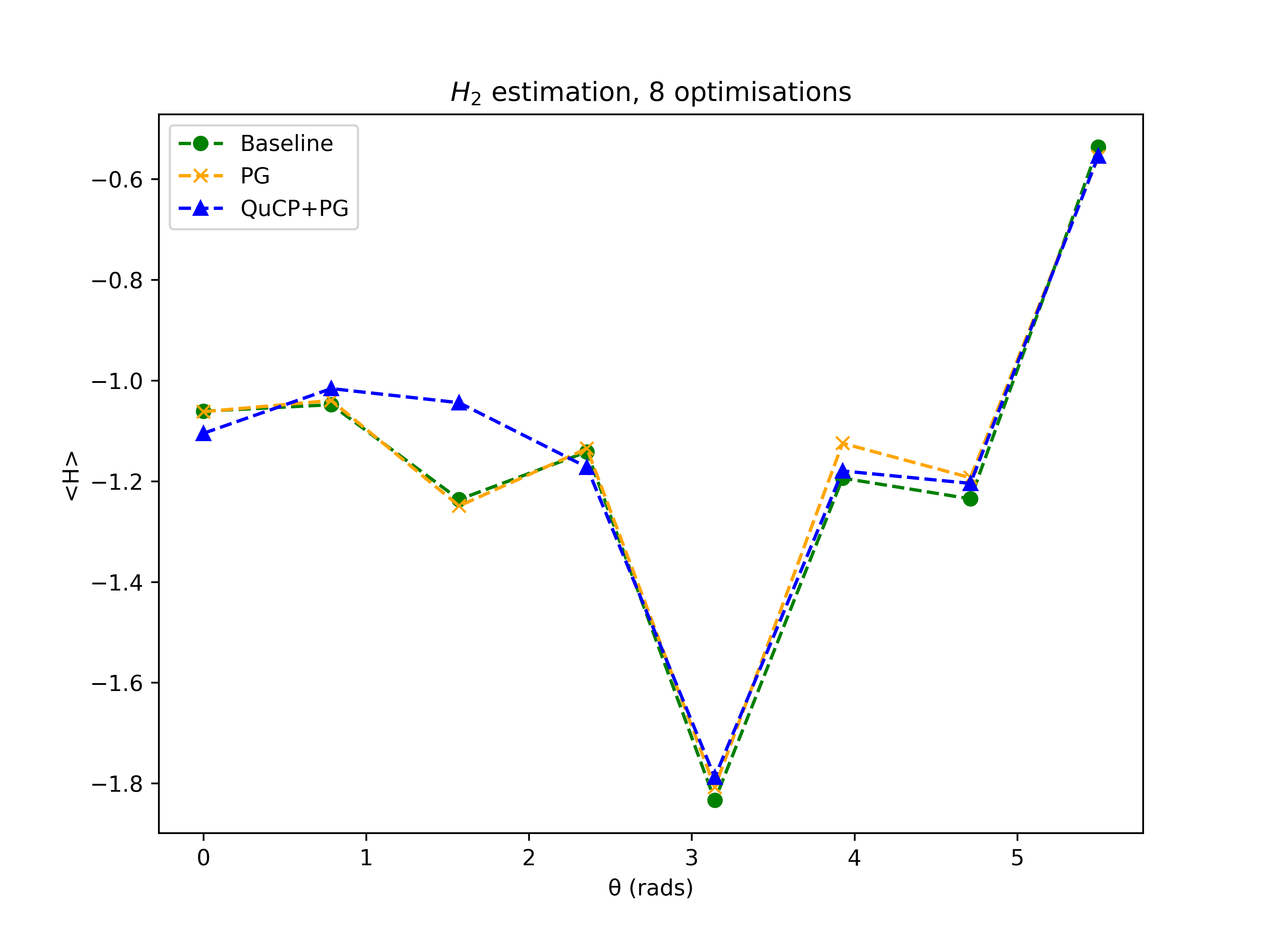}
\end{subfigure}
\hspace{0.5cm}
\begin{subfigure}{0.3\linewidth}
\centering
\caption{}
\includegraphics[scale=0.24]{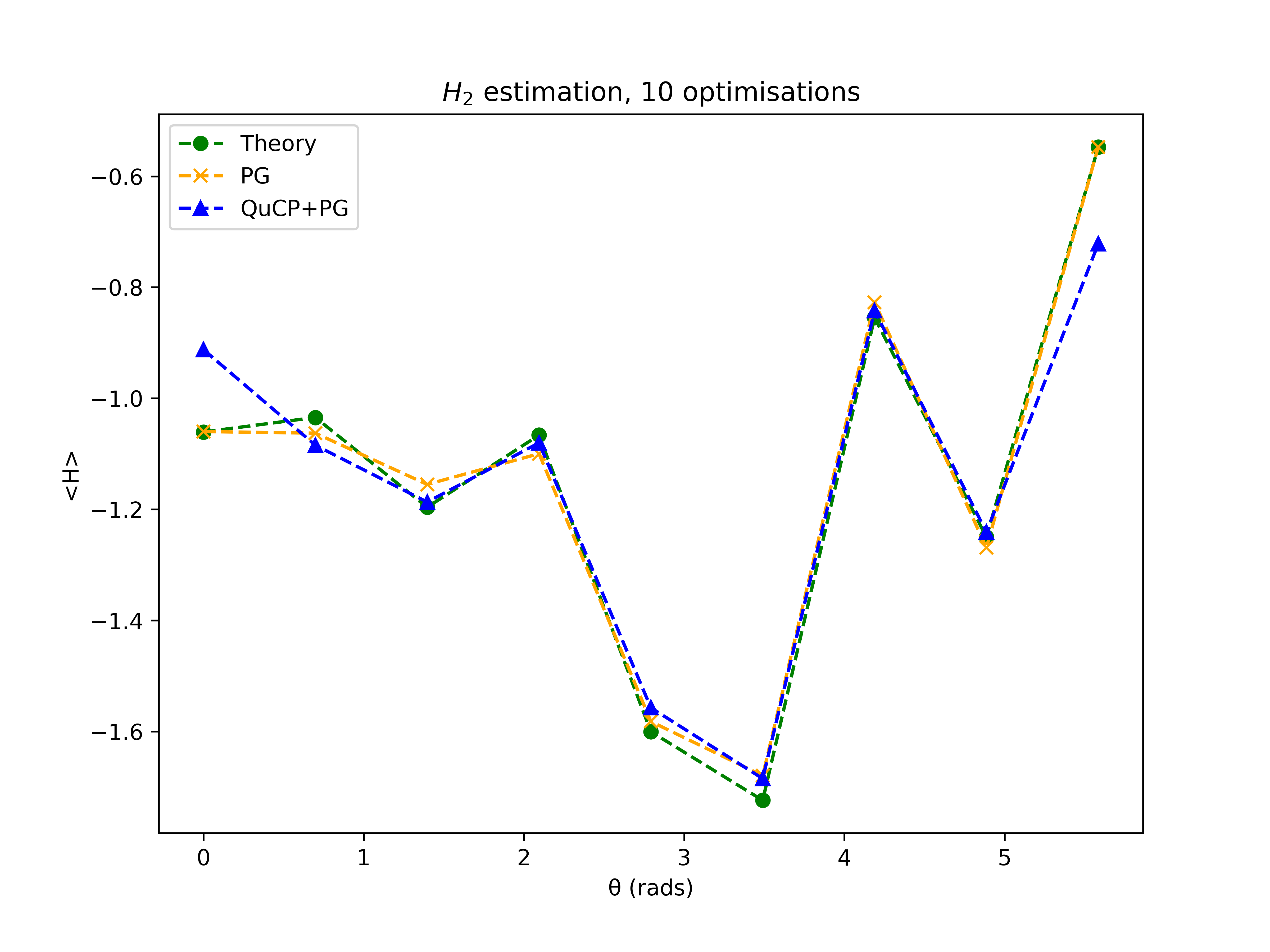}
\end{subfigure}
\hspace{0.5cm}
\begin{subfigure}{0.3\linewidth}
\centering
\caption{}
\includegraphics[scale=0.24]{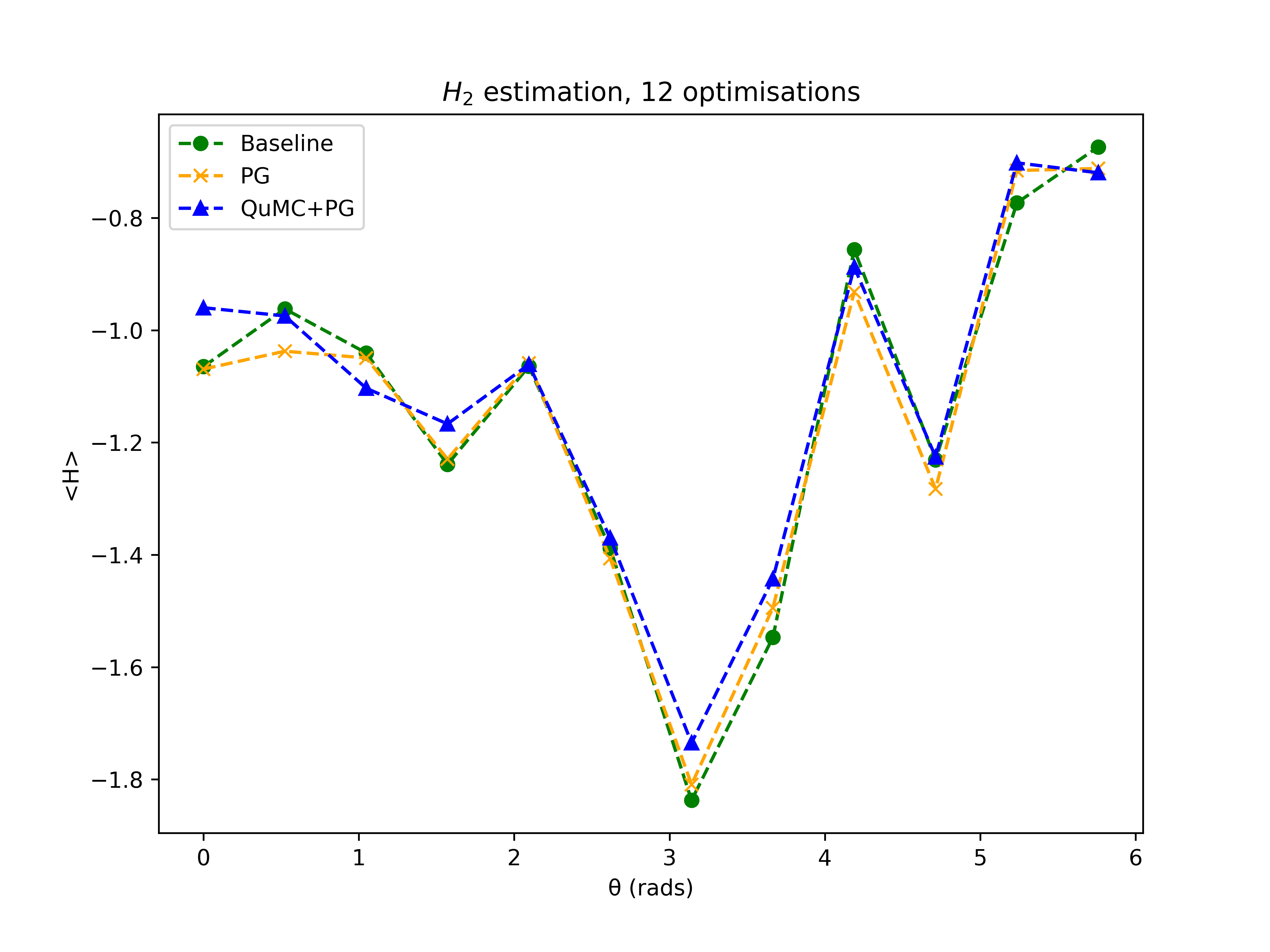}
\end{subfigure}
\caption{The estimation of the ground state energy of $H_2$ under PG and QuCP+PG. (a) 8 optimizations with 16 measurements. (b) 10 optimizations with 20 measurements. (c) 12 optimizations with 24 measurements. $n_c$ is the number of simultaneous circuit.}
\label{figure:h2}
\end{figure*}

We construct a heuristic ansatz state~\cite{kandala2017hardware} composed of two repetitions. Each repetition layer has a set of $R_yR_z$ gates on each qubit, and each qubit is entangled with the others. Overall, we have 12 parameters for single-qubit rotations and two \texttt{CNOTs} for entanglers. For the sake of simplicity, we set the same value for these parameters each time and regard them as one parameter. We choose 8, 10, and 12 parameters, which correspond to 16, 20, and 24 measurement circuits using the Pauli operator grouping (labeled as PG) simultaneous measurement method. We apply our QuCP method to PG to execute these circuits simultaneously and compare the independent process (PG) with the parallel process (QuCP + PG). Fig.~\ref{figure:h2} shows the results, and we use the result calculated by the simulator as the baseline. We pick the minimum value from the results as the ground state energy estimation to calculate the error rate compared with the baseline ($\Delta{E}$\_base). Moreover, we use the calculation of Scipy's eigensolver as the theory result and check the error rate when comparing the obtained result with the theory result ($\Delta{E}$\_theory). The information of error rates and hardware throughput of these three experiments is shown in Table~\ref{table:h2}. The hardware throughput can be up to 73.8\% with an error rate of less than 10\%. Such improvement of hardware throughput is because of the small size of the ansatz circuit with shallow depth.

\subsection{Parallel Circuit Execution and Error Mitigation}

As quantum error correction (QEC) requires a huge overhead of qubits to implement, an alternative scheme named quantum error mitigation (QEM) was proposed for error suppression on NISQ devices. There are many different error-mitigation techniques, including zero-noise extrapolation~\cite{li2017efficient}, dynamical decoupling~\cite{souza2012robust}, measurement error mitigation~\cite{bravyi2021mitigating}, etc. Among them, the zero-noise extrapolation (ZNE) method is the simplest but powerful technique that is based on error extrapolation.

ZNE was introduced in~\cite{li2017efficient}. The basic idea is to first execute the circuit in different noise levels and then extrapolate an estimated error-free value. It can be implemented in two steps: (1) Noise-scaling. (2) Extrapolation.


A digital ZNE approach was proposed in~\cite{giurgica2020digital} to scale noises by increasing the number of gates or circuit depth. A list of folded circuits with different circuit depths is generated, and we calculate their expectation values. This method only requires the programmer's gate-level access to the processor.
There are several methods for error extrapolation, such as polynomial extrapolation, linear extrapolation, and Richardson extrapolation, etc. 
However, ZNE approach introduces an overhead of executing one circuit multiple times with various depths to extrapolate the noise-free expectation value. Here, we demonstrate how to reduce this overhead by applying parallel circuit execution to the digital ZNE approach.

In our experiment, we first use \texttt{fold\_gates\_at\_random} method from Mitiq package~\cite{larose2020mitiq}. It selects gates randomly and folds them to modify the circuit depth that represents different noise levels. A list of folded circuits can be generated based on scale factors. Then, we execute these circuits simultaneously on IBM Q 65 Manhattan using the QuCP approach, and we can obtain the expectation values corresponding to different noise levels. Finally, we perform various extrapolation methods integrated in Mitiq, including \texttt{LinearFactory}, \texttt{PolyFactory}, and \texttt{RichardsonFactory}. These methods are used to calculate the estimated error-free result. One of the limitations of ZNE is that the extrapolation methods are sensitive to noises, so that the extrapolated values are strongly dependent on the selected extrapolation method. Therefore, we only show the best estimated result among these methods, which is the result that is closest to the ideal result calculated by the simulator. 

\begin{figure}[t]
\centering
\includegraphics[scale=0.35]{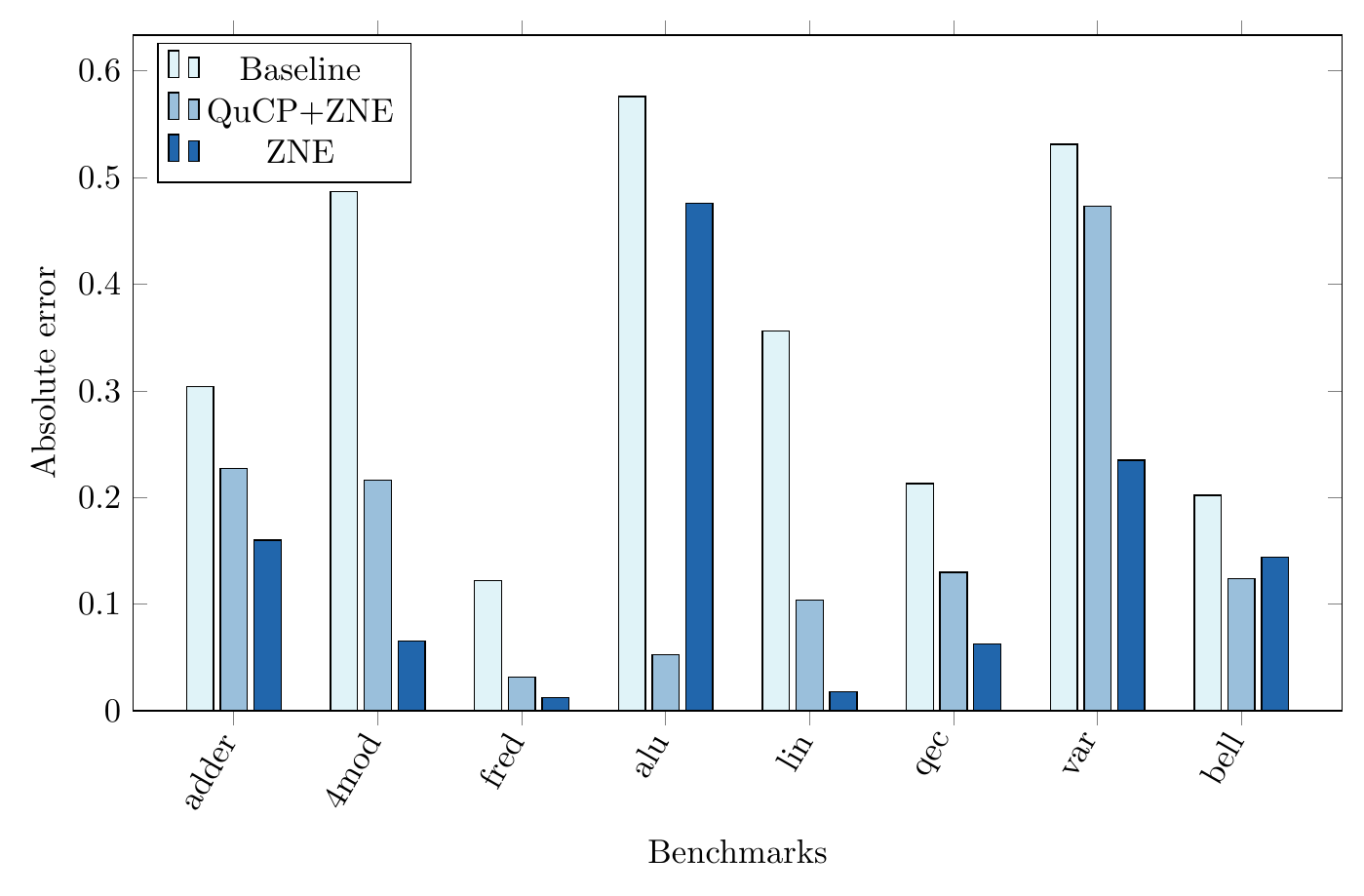}
\caption{Comparison of error rate without mitigation and with mitigation by applying QuCP multi-programming to ZNE method.}
\label{fig:zne_result}
\end{figure}

We generate four folded circuits with scale factors from 1 to 2.5 with step 0.5. Three processes are included for comparison: (1) Execute the independent circuit on the best partition selected by the QuCP method without the ZNE method (labeled as Baseline). (2) Execute the folded circuits simultaneously using QuCP to perform the ZNE method (labeled as QuCP+ZNE). (3) Execute the folded circuits independently to perform the ZNE method (labeled as ZNE). The experimental results are shown in Fig~\ref{fig:zne_result}. The absolute error is represented by the difference between the ideal expectation value calculated by the simulator and the obtained expectation value. 

According to the results, baseline always has the largest error rate due to lack of mitigation technique. In most cases, ZNE gives the lowest error rate but requires multiple circuit executions. Whereas using parallel circuit execution technique (QuCP+ZNE), the error rate can be decreased significantly compared to the baseline with the same number of circuit executions. Also, the improvement of the hardware throughput and the reduction of overall runtime is three times. On average, the error rate is reduced by 2x, and in the best case (benchmark \texttt{alu-v0\_27}), the error rate is reduced by 11x. Even though ZNE method was designed to scale the noise levels of the same occupied qubits, however, the errors can still be mitigated significantly by enlarging the circuit depth on different qubit partitions. It reveals some underlying similarities of the errors between different qubits which is interesting to explore in the future work.

\section{Conclusion}
As the size of quantum chips grows and the demand for their accessibility increases, how to efficiently use the hardware resources is becoming a concern. The parallel circuit execution mechanism has been introduced to improve the hardware throughput and reduce the task total runtime by enabling to execute multiple circuits simultaneously. In this article, we explore the parallel circuit execution technique on NISQ hardware. We first compare the state-of-the-art methods and discuss their short-comes and the impact of different factors. Second, we propose a crosstalk-aware parallel workload execution method without the overhead of crosstalk characterization. We also evaluate the NISQ hardware limitation of performing parallel circuit execution. The experiments of investigating parallel circuit execution on VQE and ZNE error mitigation method demonstrate how it can be useful for NISQ computing. It is a key enabler for quantum algorithms requiring parallel sub-problem executions, especially in NISQ era. 

\bibliography{bibliography}{}
\bibliographystyle{plain}

\end{document}